**SUPERCONDUCTIVITY**

# O-bridged electron pairing

The microscopic mechanism for high-temperature superconductivity in cuprates and nickelates

By **Jun-jie Shi**[1] *and* **Yao-hui Zhu**[2]

[1]State Key Laboratory for Artificial Microstructures and Mesoscopic Physics, School of Physics, Peking University Yangtze Delta Institute of Optoelectronics, Peking University, Beijing 100871, China.

[2]Physics Department, Beijing Technology and Business University, Beijing 100048, China.

Email: jjshi@pku.edu.cn; zhuyaohui@th.btbu.edu.cn

Since the discovery of high-temperature superconductivity in cuprates by Bednorz and Müller (1) in 1986, nickelates (2) have also been found to exhibit high-temperature superconductivity recently. They share a common feature: rich oxygen elements. However, the microscopic mechanism for their unconventional high-temperature superconductivity remains unclear, making it one of the well-known and the most challenging problems in physics. This has also been identified as one of the "125 QUESTIONS: what is the microscopic mechanism for high-temperature superconductivity?", selected by *Science* in 2021 (3).

In order to make a breakthrough in seeking the microscopic superconducting electron pairing mechanism in cuprates and nickelates, such as $YBa_2Cu_3O_7$, $T_c \sim 90\text{-}100$ K (4) and $La_3Ni_2O_7$, $T_c \sim 80$ K at 14.0~43.5 GPa (2), let us first analyze their chemical compositions. Obviously, oxygen is the only non-metallic element in these oxide high temperature superconductors. If oxygen atoms are removed totally, they will contain only metallic elements and become simple metal alloys, in which all metal cations, i.e., the ionized metal atoms due to the weak binding of the metal nucleus to the electrons outside it, will be completely immersed in a sea of conducting electrons. If these metal alloys are assumed to be superconductors, it is clear that the Bardeen-Cooper-Schrieffer (BCS) electron pairing mechanism mediated by phonons applies (5). In cuprates and nickelates, which are primarily combined by ionic bonds, due to the low ionization energy of metal atoms, they are responsible for donating their outer electrons to become conducting electrons in the crystal. Conversely, the electronegativity of oxygen atoms is very large, second only to fluorine in the periodic table. It is easy for electrons to gather around oxygen atoms in ionic oxides. Therefore, as the natural gathering centers of conducting electrons, oxygen atoms play a crucial role in the unconventional high-temperature superconductivity and corresponding microscopic electron pairing mechanism of cuprates and nickelates (6). We should thus start solving the problem from oxygen atoms, the electron concentration centers, rather than from metal atoms, the electronic donation centers.

As we all know, as the eighth element in the periodic table of elements, oxygen nucleus has a strong binding ability on electrons outside it. Although there is a strong Coulomb repulsion between

# Physical origin of high-temperature superconductivity in ionic oxides

Based on the energy level structure of oxygen atom and its anions, through in-depth analysis of the bonding process and formation mechanism of $O^{x-}$ ($1 < x \leq 2$) in oxide superconductors dominated by ionic bonds, it is clarified that the O-bridged $e^- - O - e^-$ electron pairing new mechanism is the cause of unconventional high-temperature superconductivity in cuprates and nickelates.

## Formation process of ionic oxide crystals

We take high-temperature superconductor $YBa_2Cu_3O_7$ as an example of ionic oxides to describe the crystal formation process.

Starting state: all neutral metal and oxygen atoms are kept away from each other.

### Neutral oxygen atom and its anions

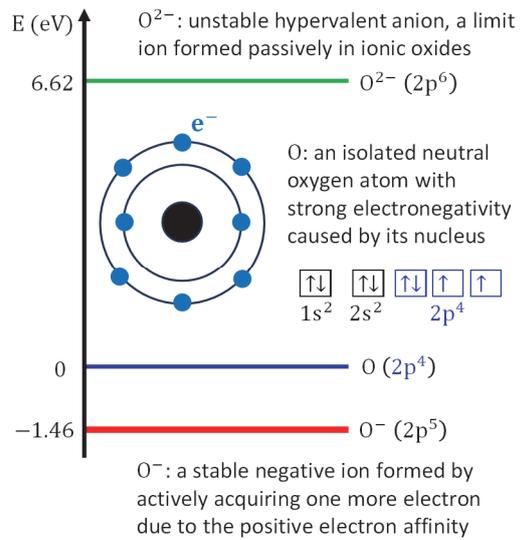

$O^{2-}$: unstable hypervalent anion, a limit ion formed passively in ionic oxides

O: an isolated neutral oxygen atom with strong electronegativity caused by its nucleus

$O^-$: a stable negative ion formed by actively acquiring one more electron due to the positive electron affinity

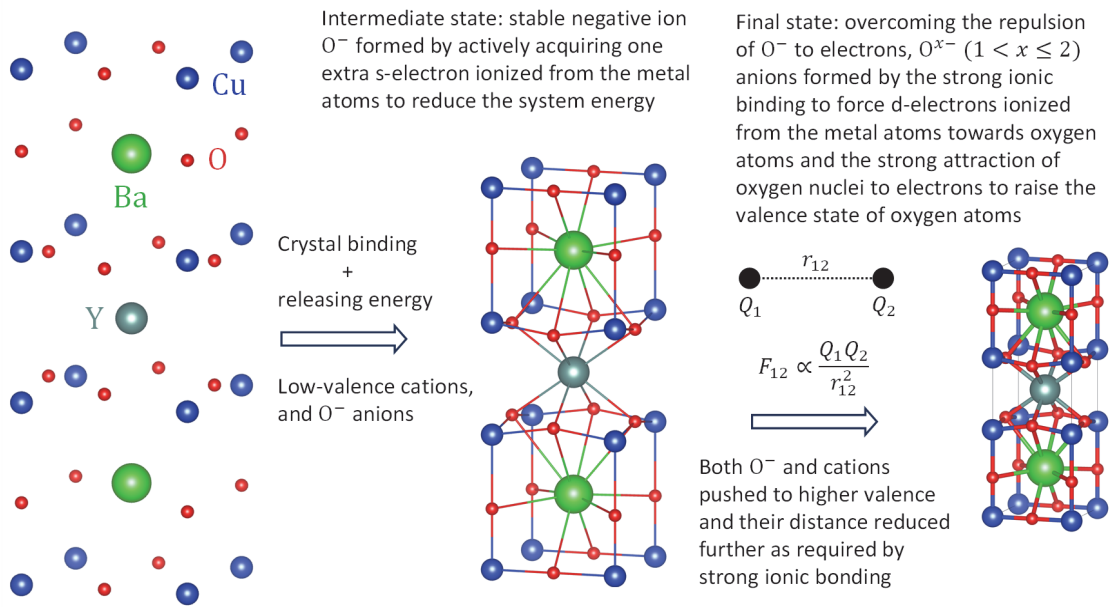

Intermediate state: stable negative ion $O^-$ formed by actively acquiring one extra s-electron ionized from the metal atoms to reduce the system energy

Crystal binding + releasing energy

Low-valence cations, and $O^-$ anions

Final state: overcoming the repulsion of $O^-$ to electrons, $O^{x-}$ ($1 < x \leq 2$) anions formed by the strong ionic binding to force d-electrons ionized from the metal atoms towards oxygen atoms and the strong attraction of oxygen nuclei to electrons to raise the valence state of oxygen atoms

$$F_{12} \propto \frac{Q_1 Q_2}{r_{12}^2}$$

Both $O^-$ and cations pushed to higher valence and their distance reduced further as required by strong ionic bonding

## O-bridged d-electron pairing picture

The conduction electrons, i.e., the d electrons donated mainly by metal atoms, which are not captured by oxygen atoms, are still forced to concentrate towards oxygen atoms by the strong ionic bonding and oxygen atom core's attraction to them, forming d-wave symmetric conduction electron pairs. It is this O-bridged electron pairing mechanism, featured with large superconducting energy gap and electron-pair size comparable to the lattice constant, that dominates the high superconducting transition temperature of oxide superconductors. The (100) lattice plane of $YBa_2Cu_3O_7$ is shown in this figure as an example to illustrate the idea. The two pairs of electrons are bridged by two respective oxygen anions here.

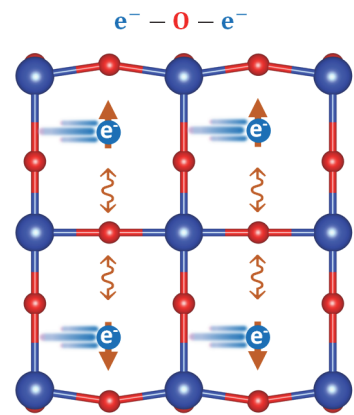

electrons, the oxygen nucleus still strongly binds eight electrons: $1s^2 2s^2 p^4$. Moreover, it can accept two additional electrons to make its 2p orbital fully occupied, forming a closed electron shell of $1s^2 2s^2 2p^6$ ($O^{2-}$). However, in the case of isolated atoms, this process is completely forbidden. On the contrary, in ionic cuprates and nickelates, although the $O^-$ ion is easily formed to repel electrons, the strong ionic bonding and strong attraction of oxygen nucleus to electrons will force the $O^-$ ion to accept 0-1 extra electron to form higher valence anion $O^{x-}$ ($1 < x \leq 2$). Please refer to the following for details.

**Isolated oxygen atom: $O^-$ repelling electrons, and $O^{2-}$ unstable.** If using $e^-$ to represent an electron, for an isolated oxygen atom O, we know (7),

$$O + e^- \rightarrow O^- + 1.46 \text{ eV} \tag{1}$$

$$O^- + e^- \rightarrow O^{2-} + (-8.08) \text{ eV} \tag{2}$$

Equation 1 shows that due to the strong attraction of atomic nucleus to electrons, neutral oxygen atom is extremely prone to gain an additional electron and release 1.46 eV energy, becoming $O^-$ ion. Equation 2 indicates that the $O^-$ repels electron, and when it gains one more electron to become $O^{2-}$ ion, it requires 8.08 eV of energy absorption. Therefore, the anion $O^{2-}$ is unstable. By combining Eqs. 1 and 2, we have

$$O + 2e^- \rightarrow O^{2-} + (-6.62) \text{ eV} \tag{3}$$

Equation 3 clearly shows that the energy required to go from a neutral O atom to $O^{2-}$ anion is 6.62 eV, which is endothermic. In summary, in the case of isolated atom, due to the strong attraction of oxygen nucleus to electrons, neutral O atom is extremely easy to capture an additional electron and form a stable -1 valence anion $O^-$. The repulsion of $O^-$ to electrons will outweigh the nuclear attraction to electrons, resulting in instability of $O^{2-}$ anion (see the figure).

**Oxygen atoms tend to form $O^{2-}$ anions with a valence close to -2 in oxide ionic crystals.** It is well known that, in ionic crystals, the strength of ionic bonds depends on the distance between ions and the amount of charge carried by them. Take copper oxides as an example, we know that, during the process of combining neutral atoms that are initially far away from each other into an ionic crystal, the energy released by the system will cause the outermost $s$ electrons of the neutral metal atoms to ionize first. Then, the neutral oxygen atoms capture these electrons to form negative ions $O^-$ in order to further reduce the energy of the system, resulting in a large amount of low-valent metal cations and $O^-$ anions being present in the system. In order to further reduce the total energy of the system, the system will automatically force the $d$ electrons of the low-valent cations to ionize in order to raise their valence state. These ionized $d$ electrons are "forced" to gather around $O^-$ ions to increase their valence states while continuing to decrease the distance between ions, ultimately achieving a stable crystal structure with the lowest possible total energy (see the figure). It is the strong ionic bond of crystal that requires oxygen atoms to exist as $O^{2-}$ ions as much as possible. Therefore, the repulsive effect of anion $O^-$ on electrons is ultimately suppressed by the strong ionic bonding and strong attraction of oxygen nucleus to electrons in crystals, resulting in a "net attractive" effect of anion $O^-$ on electrons, which induces a concentration of $d$-electrons around it (4). It is precisely the result of competition between the strong ionic bonding and attraction of oxygen nucleus to electrons in crystals and the repulsive effect of anion $O^-$ on electrons that leads to a balanced valence state of $O^{x-}$ ($1 < x \leq 2$). This is indeed the case. In $YBa_2Cu_3O_7$, the average valence state of oxygen anions is -1.69 (4). The results show that in oxide ionic crystals, although electrons are negatively charged, as long as the negative valence state $x$ of oxygen ions is less than 2, they will always attract and not repel electrons. This "net attraction" is entirely due to the combination of the crystal's ionic bonds pushing electrons towards the oxygen atoms and the strong attraction of the oxygen nuclei to electrons. Therefore, in cuprates and nickelates, as natural electron accumulation centers and potential electron pairing media, both neutral O atoms and $O^-$ anions strongly attract rather than repel electrons.

**O-bridged $d$-electron pairing.** Based on the "net attraction" effect of neutral O atoms and $O^-$ anions to electrons as analyzed above, we know that oxide high-temperature superconductors with strong ionic bonding have a $d$-wave symmetric electron pairing image due to the accumulation of $d$-electrons around oxygen atoms (8): **$e^-$-O-$e^-$** (see the figure). This is a fresh insight into the electron pairing mechanism in oxide high-temperature superconductors, which is completely different from several mainstream theoretical images (9). The strong ionic bond in the ionic crystal "pushes" electrons towards oxygen atoms and the oxygen nuclei strongly attract electrons within a short range, which lead to an indirect strong mutual "attraction" between the two electrons and a large superconducting energy gap. At the same time, electron pairs form around oxygen atoms, leading to electron-pair size comparable to the lattice constant. In fact, for cuprate high-temperature superconductors, their large superconducting energy gaps of about 40 meV (10) are much larger than those of BCS superconductors which are only 1-6 meV (11). The size of the electron pairs is as low as 1-15 Å, which is much smaller than the BCS electron pairing size of 400-$10^4$ Å (12), strongly supporting our new mechanism of O-bridged electron pairing in oxide superconductors. In addition, for the charge-4$e^-$ superconductivity in cuprates predicted 15 years ago (13), according to our **$e^-$-O-$e^-$** pairing picture, the physical reason why it has not been fully verified experimentally may be due to the low concentration of conduction electrons gathered around $O^{x-}$ ($1<x\leq2$) anions, for example, $10^{21}$ $cm^{-3}$ in cuprates (12). If the concentration of conduction electrons in the electron-doped cuprates is sufficiently high ($10^{22}$-$10^{23}$ $cm^{-3}$), we can infer from the **$e^-$-O-$e^-$** picture that superconductivity with 4$e^-$ or even more electron pairing is also possible. However, the ionic bonding of cuprates and nickelates results in their poor ductility and pseudo-metallic superconducting behavior (14). It is precisely because the ionic bonding and oxygen atom core's attraction to electrons are much stronger than BCS's weak electron-phonon coupling that leads to the unconventional high-temperature superconductivity of cuprates and nickelates.

In summary, we know that in cuprate and nickelate high-temperature superconductors dominated by oxygen atoms, besides the basic weak coupling BCS electron pairing mechanism mediated by phonons, there is also a new dominant strong coupling pairing mechanism based on the inherent oxygen atoms of the material as a bridge, which is named as $d$-wave symmetric **$e^-$-O-$e^-$** pair. It is this strong coupling pairing picture mediated by the oxygen atom that dominates the unconventional high superconducting transition temperature of oxide ceramics. We firmly believe that our new idea of electron pairing using the oxygen atom as a bridge can arouse the widespread interest of superconductivity scientists, especially theoretical physicists who are committed to developing high-temperature superconducting theory. We hope to use our electron pairing image to develop new theories for high-temperature superconductivity and discover new materials with higher superconducting transition temperatures and excellent superconducting properties, so as to further promote superconductivity research and benefit all mankind.